# Giant Modulation of Refractive Index from Picoscale Atomic Displacements

*Boyang Zhao, Guodong Ren, Hongyan Mei, Vincent C. Wu, Shantanu Singh, Gwan-Yeong Jung, Huandong Chen, Raynald Giovine, Shanyuan Niu, Arashdeep S. Thind, Jad Salman, Nick S. Settineri, Bryan C. Chakoumakos, Michael E. Manley, Raphael P. Hermann, Andrew R. Lupini, Miaofang Chi, Jordan A. Hachtel, Arkadiy Simonov, Simon J. Teat, Raphaële J. Clément, Mikhail A. Kats, Jayakanth Ravichandran[\*], and Rohan Mishra[\*]*

Boyang Zhao, Shantanu Singh, Huandong Chen, Shanyuan Niu, Jayakanth Ravichandran
Mork Family Department of Chemical Engineering and Materials Science, University of Southern California, Los Angeles, CA, 90089, USA
E-mail: j.ravichandran@usc.edu

Guodong Ren, Arashdeep S. Thind, Rohan Mishra
Institute of Materials Science and Engineering, Washington University in St. Louis, St. Louis, MO, 63130, USA
E-mail: rmishra@wustl.edu

Hongyan Mei, Jad Salman, Mikhail A. Kats.
Department of Electrical and Computer Engineering, University of Wisconsin–Madison, Madison, WI, 53706, USA

Vincent C. Wu, Raynald Giovine, Raphaële J. Clément
Materials Department and Materials Research Laboratory, University of California, Santa Barbara, CA 93106, USA

Shanyuan Niu
Present Address: College of Engineering and Applied Sciences, National Laboratory of Solid State Microstructures, Nanjing University, Nanjing, 210093, China

Gwan-Yeong Jung, Rohan Mishra
Department of Mechanical Engineering and Materials Science, Washington University in St. Louis, St. Louis, MO, 63130, USA

Nick S. Settineri, Simon J. Teat

Advanced Light Source, Lawrence Berkeley National Laboratory, Berkeley, CA, 94720, USA

Bryan C. Chakoumakos

Neutron Scattering Division, Oak Ridge National Laboratory, Oak Ridge, TN, 37831, USA

Michael E. Manley, Raphael P. Hermann

Materials Science and Technology Division, Oak Ridge National Laboratory, Oak Ridge, TN, 37831, USA

Andrew R. Lupini, Miaofang Chi, Jordan A. Hachtel

Center for Nanophase Materials Sciences, Oak Ridge National Laboratory, Oak Ridge, TN, 37831, USA

Arkadiy Simonov

Department of Materials, ETH Zurich, Vladimir-Prelog-Weg 1-5/10, 8093 Zürich, Switzerland.

Jayakanth Ravichandran

Ming Hsieh Department of Electrical Engineering, University of Southern California, Los Angeles, CA, 90089, USA

Core Center of Excellence in Nano Imaging, University of Southern California, Los Angeles, CA, 90089, USA

**Abstract**

Structural disorder has been shown to enhance and modulate magnetic, electrical, dipolar, electrochemical, and mechanical properties of materials. However, the possibility of obtaining novel optical and optoelectronic properties from structural disorder remains an open question. Here, we show unambiguous evidence of disorder — in the form of anisotropic, picoscale atomic displacements — modulating the refractive index tensor and resulting in the giant optical anisotropy observed in $BaTiS_3$, a quasi-one-dimensional hexagonal chalcogenide. Single crystal X-ray diffraction studies reveal the presence of antipolar displacements of Ti atoms within adjacent $TiS_6$ chains along the *c*-axis, and three-fold degenerate Ti displacements in the *a-b* plane. $^{47/49}$Ti solid-state NMR provides additional evidence for those Ti displacements in the form of a three-horned NMR lineshape resulting from a low symmetry local environment around Ti atoms. We used scanning transmission electron microscopy to directly observe the globally disordered Ti *a-b* plane displacements and find them to be ordered locally over a few unit cells. First-principles calculations show that the Ti *a-b* plane displacements selectively reduce the refractive index along the *ab*-plane, while having minimal impact on the refractive index along the chain direction, thus resulting in a giant enhancement in the optical anisotropy. By showing a strong connection between structural disorder with picoscale displacements and the optical response in $BaTiS_3$, this study opens a pathway for designing optical materials with high refractive index and functionalities such as large optical anisotropy and nonlinearity.

# 1. Introduction

Crystalline matter is defined by the presence of periodic order, although *real* crystalline materials possess disorders in various forms across multiple length scales[1]. While disorder can be detrimental to physical properties, it can also lead to emergent physical properties and influence phase transitions[2]. The latter is especially true for correlated disorder, which signifies the presence of short-range correlations between some structural features in an otherwise disordered structure[3]. For example, correlated disorder leads to large electro-mechanical coupling in ferroelectric alloys[4,5], spin frustration and glassy behavior in magnets[6], colossal magneto-resistive effects in manganites[7], unconventional metal-insulator transitions in disordered metals and semiconductors[8], enhanced superconducting phase fluctuations[9], and enhanced Li-ion conductivity in oxide alloys[10].

Structural disorder has a dramatic effect on the static or zero-frequency dielectric response of materials. For instance, relaxor ferroelectric alloys exhibit static dielectric constants that are orders of magnitude higher than unary ferroelectric compounds, with the correlated disorder being the origin of the enhanced response[5]. Similar enhancements in the high-frequency optical properties of materials, such as their refractive index ($n$), have not been observed with the structural disorder. Furthermore, given the strong correlation between the refractive index of a material and its nonlinear optical response[11], new approaches to enhance the refractive index could also result in additional functionalities such as large optical anisotropy and nonlinearities. Here, we show that structural disorder results in a large change in the refractive index along specific crystallographic directions in a quasi-one-dimensional (1D) hexagonal chalcogenide, $BaTiS_3$, and leads to giant optical anisotropy that has been reported previously[12] but remained poorly understood[13,14]. These results suggest that atomic displacements, even when disordered, can be used as an additional degree of freedom[15] to design high refractive index optical materials for communication and sensing applications.

Optical anisotropy is characterized by birefringence ($\Delta n$) and dichroism ($\Delta \kappa$), which are, respectively, the differences in the real ($n$) and imaginary ($\kappa$) parts of the complex refractive index between two crystallographic directions. Organic materials demonstrate stereoisomerism[16], and one can leverage large dipolar rearrangements in these materials to achieve dramatic changes in anisotropic optical properties such as optical activity, birefringence and dichroism, and nonlinear optical properties[17]. In crystalline materials, an anisotropic crystal structure and favorable polarizability of the component elements are desirable for optical anisotropy[18]. BaTiS$_3$ is a quasi-1D semiconductor that has been measured to have giant optical anisotropy (Figure 1a)[12,19]. The infrared (IR) birefringence of BaTiS$_3$ is as high as ~0.76, which is one of the highest birefringence values achieved by a crystalline material in the transparent regime, with only a few, recently reported, compounds showing higher values in the IR spectral range[20,21].

Single crystals of BaTiS$_3$ were reported to crystallize in *P6$_3$mc* space group[12,22] (unit cell visualized in Figure 1c for reference), whose face-shared TiS$_6$-octahedra form quasi-1D chains along the *c*-axis. The measured giant optical anisotropy was qualitatively explained in terms of the anisotropic distribution of elements with large differences in electronic polarizability[12]; however, the theoretically predicted and experimentally observed values of optical anisotropy were not in agreement[13] First-principles, density-functional theory (DFT) calculations of $\Delta n$ for BaTiS$_3$ with *P6$_3$mc* space group show that it is moderately birefringent (Figure 1a and S2a)[14], on the same level as other birefringent crystals such as rutile (~0.25)[23], but much lower than the experimentally measured values for BaTiS$_3$ (~0.76)[12]. Furthermore, the theoretically predicted dichroism shows an incorrect spectral dependence compared to the measured spectrum (Figure S2b). Using different exchange-correlation functionals within DFT does not ameliorate the situation (SI Section I).

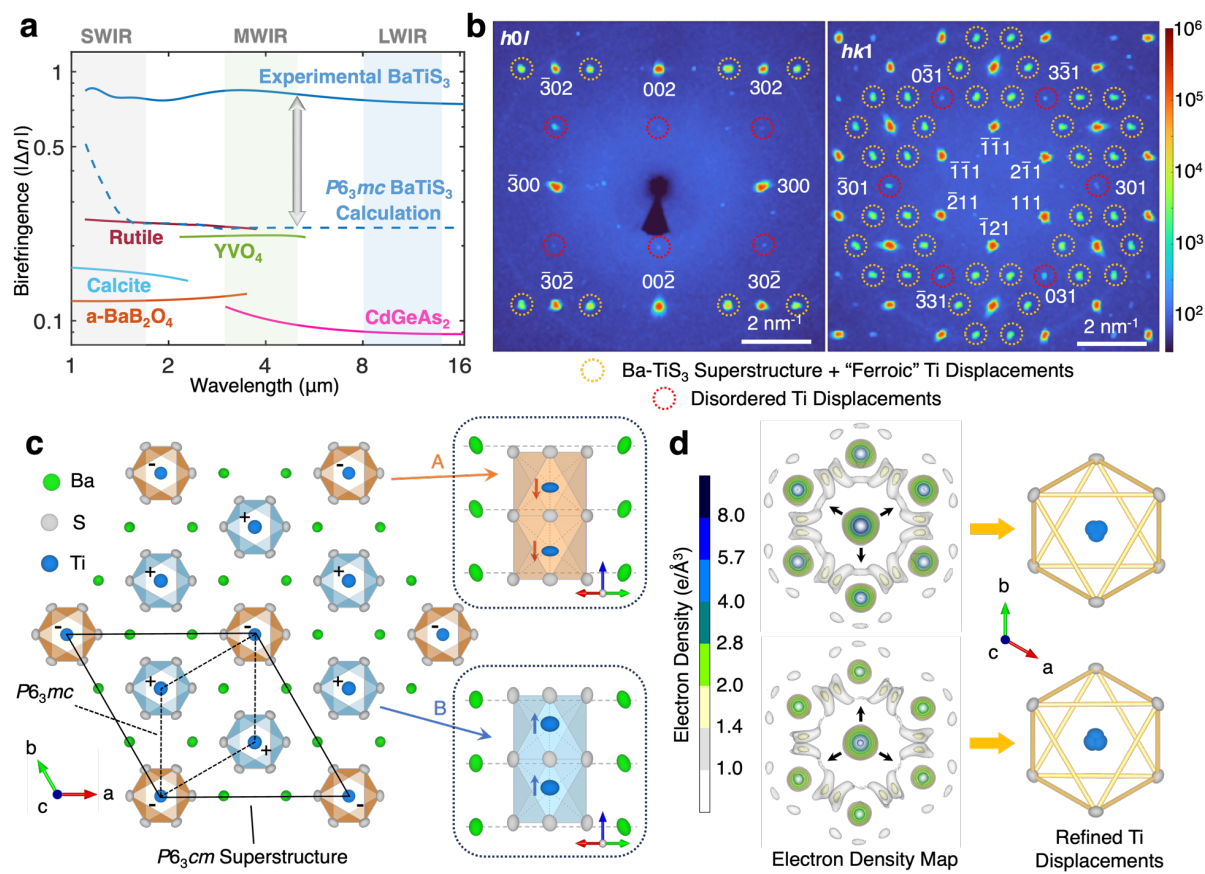

**Figure 1. Single crystal X-ray diffraction of BaTiS$_3$.** [**a**] Comparison of the birefringence of BaTiS$_3$, which was experimentally measured[12] (blue solid line) and theoretically calculated from first-principles using the previously reported $P6_3mc$ space group (dashed blue line) against other birefringent crystals. The disagreement in the birefringence of BaTiS$_3$ between experiment and theory is evident. [**b**] Reciprocal-space precession maps for $h0l$- and $hk1$-type reflections from synchrotron single-crystal diffraction. Satellite reflections caused by a Ba-TiS$_3$ superstructure are highlighted with yellow dotted circles while weak reflections denoting disordered displacements are shown with red dotted circles. [**c**] Schematic of the BaTiS$_3$ crystal structure projected onto the $a$-$b$ plane from the refined diffraction results. The ellipsoids used to show Ba, Ti, and S atoms reflect their refined atomic displacement parameters (ADPs). The trimerized $\sqrt{3} \times \sqrt{3} \times 1$ superstructure of TiS$_6$ chains leads to a lower symmetry $P6_3cm$ space group, which is the result of antiparallel displacements along the $c$-axis. The insets on the right visualize the positions of TiS$_6$ chains A (brown) and B (blue) with respect to adjacent Ba atoms. Chain A displaces down while chain B displaces up from a "fixed" Ba lattice. Moreover, Ti atoms also move away from the centroid of the S$_6$ octahedron antiparallelly between chain A and chain B. [**d**] Electron density maps, and displacement analysis for Ti01 (in chain A) and Ti02 (in chain B) along the $a$-$b$ plane. Electron density below 1 e/Å$^3$ is comparable with the noise level and thus whitened. An anharmonic core electron distribution for Ti reveals disordered non-thermal Ti *a-b* plane displacements towards S atoms.

## 2. Superstructural Lattice Distortions along the *c*-axis

Recent structural characterization using neutron pair distribution function analysis on polycrystalline BaTiS$_3$ in the *P*6$_3$*mc* space group revealed an anomalously large temperature dependence of the anisotropic atomic displacement parameters (ADPs), much beyond thermal broadening[24]. We observed symmetry-forbidden Bragg reflections in high-resolution X-ray diffraction (XRD) results on BaTiS$_3$ crystals [Figure S3a], which motivated us to revisit the experimentally reported structure of BaTiS$_3$. Synchrotron single crystal X-ray diffraction (SC-XRD) studies were performed on crystalline BaTiS$_3$ samples — that were synthesized using a previously reported method[12,19]. We observe weak Bragg reflections suggesting a lower symmetry than *P*6$_3$*mc* (See the Methods section for details of the experiments). The integrated reciprocal space structure can be best represented by *h*0*l* and *hk*1 precession maps shown in Figure 1b. Even though the previously assigned *P*6$_3$*mc* space group[3,7] captures the main reflections, we observe weaker symmetric superlattice reflections (yellow dotted circles in Figure 1b) corresponding to a $\sqrt{3} \times \sqrt{3} \times 1$ trimerized translational symmetry over the *P*6$_3$*mc* lattice. Other symmetry elements, e.g., the screw axes (6$_3$), rotation axes (3), the mirror (*m*), and glide (*c*) planes span the same orientations but in a resized and reoriented [Figure S3b-d] unit cell. By including these superlattice reflections, we could figure out the formerly unassigned lattice disorder[3,7], which was referred to as potential *P*6$_3$*mc* domain anomalies, to be refined as periodic atomic displacements in a *P*6$_3$*cm* space group of a $\sqrt{3} \times \sqrt{3} \times 1$ unit cell (for details, see Supporting Information Section VIII with Table S4, S5, S6 and S7). The resulting *P*6$_3$*cm*-BaTiS$_3$ structure projected onto the *a-b* plane is shown in Figure 1c. The previously degenerate TiS$_6$-chains are now split into two types, each occupying different positions [right insets of Figure 1c] along the *c*-axis and with antiparallel Ti off-centering along the *c*-axis. They are labeled as TiS$_6$-chain A at the edges of the unit cell and TiS$_6$-chain B inside the unit cell; each unit cell thus has one chain A and two chains B. The Ti atoms are also displaced along the *c*-axis from the S$_6$ centroid, downwards by 0.167 Å in chain A and upwards by 0.147 Å in chain B. These antiparallel off-center displacements result in a ferrielectric ordering, as opposed to the ferroelectric ordering proposed for the *P*6$_3$*mc* space group[24].

Despite the significant improvement in the refinement residual of only 1.84% with the *P*6$_3$*cm* space group (see Supporting Information Section VIII), we find systematic deviations in the intensity of the reflection spots between the measured and refined diffraction patterns. For example, we observe weak signals of certain symmetry-extinct reflections ($h0l$ or $0kl$ peaks whose $l = 2n+1$) that are highlighted by red dotted circles in Figure 1b. The existence of such reflections suggests the presence of subtle displacements that may locally distort the $P6_3cm$ symmetry. As atomic displacement parameters (ADPs) [Table S7] capture the average atomic displacements about the statistical centroid of atomic positions[25], large ADPs, well beyond the thermal vibrations (details see Supporting Information Section IV), are another sign of the presence of subtle displacements. The anomalously large ADPs along the $c$-axis ($U_{33}$) reported earlier[12,24] have already been resolved by refining the diffraction data with the $P6_3cm$ structure. Nevertheless, the ADPs of Ti along the $a$-$b$ plane ($U_{11}$ and $U_{22}$) are still larger than that of Ba and the lighter S atoms[26], suggesting potential displacements of Ti atoms within the TiS$_6$ octahedra in the $a$-$b$ plane (ellipsoidal atoms in Figure 1c and Figure 1d covers the 90% and 50% possibility region finding corresponding atoms, a result of ADPs refinement).

## 3. Symmetry-Breaking along the *a-b* Plane

To understand the nature of this weak symmetry breaking and large ADPs, we examined the difference between the observed and refined electron densities, also known as the Fourier difference map (Supporting Information Section III). The Fourier difference map shows small but definite, deviations in the partial occupancy of Ti away from the centroid of the two types of TiS$_6$ chains. This deviation is highlighted in Figure 1d using electron density maps projected onto the $a$-$b$ plane about the refined Ti A and Ti B sites. Here, we observe statistically significant, three-fold degenerate Ti atomic displacements (as compared to random displacements indistinguishable from the thermal vibrations) having an occupancy ~21% away from the centroid (Supporting Information Section III). The magnitude of such $a$-$b$ plane Ti displacements is obtained from the refined structure (details see Supporting Information Section III and Table S4, S6, S7). This structure further lowers the refinement residual ($R_1 = 1.71\%$) and the ADPs of corresponding Ti atoms [Table S7], suggesting that these displacements are relevant. We do not discern whether the $a$-$b$ plane Ti displacements are static or dynamic as they are likely to have a similar effect on the high-frequency optical response. The temperature-dependent phonon dispersion[27] of the $a$-$b$ plane off-centering mode may have implications on inelastic scattering and thermal properties.

X-ray diffraction relies on (core and valence) electron density to probe the positions of the nuclei, so we resorted to techniques directly sensitive to the nuclei to confirm these observations.

We derived the pair distribution function (PDF) from the reported neutron scattering[24], which shows an improved fit to the peak corresponding to the first nearest Ti–S distance [Figure S6e] for the *P*6$_3$*cm* structure with *a-b* plane displacements while maintaining a consistent long-range order [Figure S6c-d]. Ti displacements in BaTiS$_3$ were further examined using solid-state nuclear magnetic resonance (ssNMR) spectroscopy. ssNMR is sensitive to the short-range structure around the Ti nuclei of interest. As such, it provides a window into deviations from symmetry in the Ti coordination environment. Titanium has two NMR-active isotopes, $^{47}$Ti and $^{49}$Ti, with remarkably similar gyromagnetic ratios (*γ*) of −1.5105 and −1.51095 rad·s$^{-1}$·T$^{-1}$, resulting in very similar NMR resonant frequencies and overlapping signals. Two types of interactions dictate $^{47/49}$Ti ssNMR lineshapes: the chemical shift anisotropy (CSA) resulting from a non-spherical electron density around the $^{47/49}$Ti nucleus, and first- and second-order quadrupolar interactions between the quadrupolar moment of $^{47}$Ti (*I*=5/2) and $^{49}$Ti (*I*=7/2) nuclei and the local electric field gradient (EFG). While both CSA and quadrupolar effects are accounted for in the following analysis, we focus our discussion on the latter as they dominate and provide insights into the centrosymmetry (through the quadrupolar coupling constant, $C_Q$) and asymmetry (through the quadrupolar asymmetry, $\eta_Q$) of Ti local environments.

To address issues of low sensitivity and wide excitation ranges associated with $^{47/49}$Ti ssNMR[28], and to minimize quadrupolar broadening effects, static $^{47/49}$Ti ssNMR spectra were acquired at low temperature (270 K) and a high magnetic field (18.8 T, 800 MHz for $^1$H) using a WURST-QCPMG experiment (full acquisition details are described in the Experimental section). The $^{47/49}$Ti ssNMR spectrum obtained on BaTiS$_3$ is shown in black in Figure 2a-b, where the normally continuous spectrum is instead represented by an envelope of spikelets resulting from the QCPMG experiment. To facilitate the assignment of experimental $^{47/49}$Ti ssNMR data, first principles-guided fits of the ssNMR lineshape (red line) were performed using two structural models (methodology explained in Section VI in the Supporting Information): the undistorted BaTiS$_3$ structure (*P*6$_3$*cm* space group) obtained from SC-XRD [Figure 2a], and a BaTiS$_3$ structure with Ti *a-b* plane displacements obtained by freezing the $\Gamma_5$ distortion mode and optimized from first principles using the Vienna Ab initio Simulation Package (VASP)[29,30] [Figure 2b]. The experimental $^{47/49}$Ti ssNMR spectrum contains three main peaks, labeled "left", "center", and "right", which is inconsistent with the axially symmetric NMR properties of the two Ti local environments (Ti01 and Ti02) in the undistorted BaTiS$_3$ structure, resulting in a poor fit [Figure 2a]. In contrast, a much better fit of the spectrum is obtained using the distorted BaTiS$_3$ structural model [Figure 2b]: in this structure, the close to axially symmetric ($\eta_Q = 0.2$)

Ti01 environment has a large quadrupolar coupling constant, $C_Q$, resulting in a double-horn signal spanning the "left" and "right" peaks of the experimental spectrum, and the Ti02 environment has a very high quadrupolar asymmetry, $\eta_Q = 0.7$, leading to the "center" peak observed experimentally. These results confirm the presence of large Ti *a-b* plane displacements in the BaTiS$_3$ structure, resulting in less symmetric Ti environments.

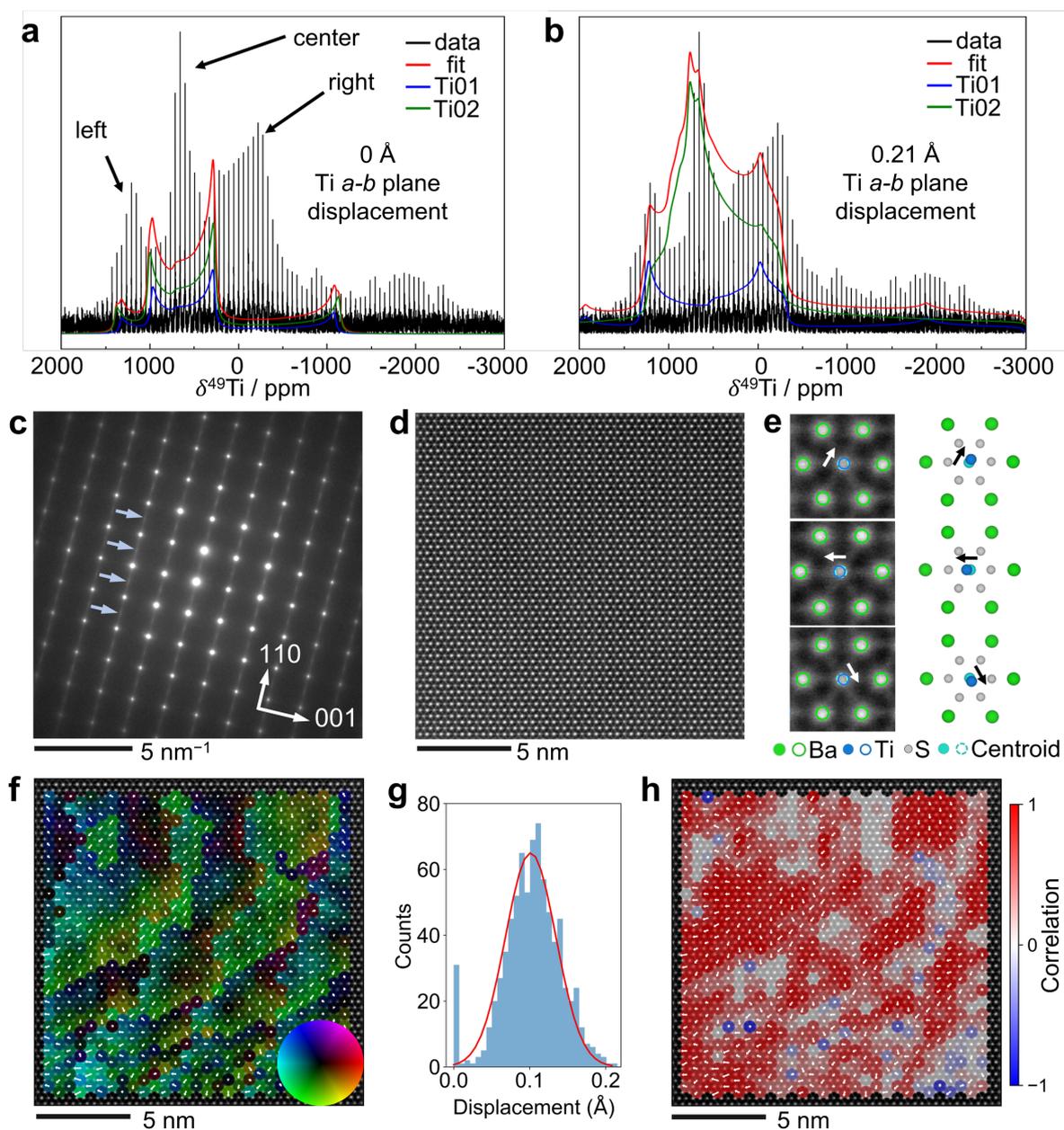

**Figure 2. Ti displacements along the *a-b* plane from NMR with electron microscopy providing direct evidence of their correlated disorder.** Static $^{47/49}$Ti solid-state NMR spectra collected on BaTiS$_3$ at 270 K and 18.8 T, overlaid with first-principles guided fits of $^{47/49}$Ti NMR spectra derived from **a**, the $P6_3cm$ BaTiS$_3$ structure obtained from SC-XRD and with no Ti *a-b* plane displacements, and **b**, the BaTiS$_3$ structure obtained from DFT calculations, and exhibiting 0.213 Å Ti *a-b* plane displacements. **c**, TEM diffraction pattern viewed along the [1$\bar{1}$0] zone axis. Diffuse streaks specifically extend along the [110] direction, while sharp Bragg peaks exist along the [001] direction. **d,** Representative HAADF-STEM image showing the atomic structure of BaTiS$_3$ along the *a-b* plane projection. **e,** Local structural motifs from the HAADF image in **d** showing Ti *a-b* plane displacement towards adjacent S atoms. Ti displacements are determined from the centroid of the surrounding hexagonal Ba sublattice. **f,** Ti displacement vector map calculated from the HAADF-STEM image of BaTiS$_3$ along [001]-zone axis overlaid onto the original HAADF image in **d**. The direction and magnitude of Ti displacements are represented by arrows. The direction is also indicated by the color wheel for easier visualization. **g,** Histogram of the magnitude of Ti *a-b* plane displacements obtained from the HAADF image in **d**. Aside from a small number of Ti atoms that do not undergo off-centering in the *a-b* plane, resembling boundaries of domains with displacements, the average displacement is ~0.11 Å. The distribution is fitted with a Gaussian. **h,** Linear correlation map of Ti *a-b* plane displacement vectors. The color map shows the correlation coefficient of nearest-neighbor displacement vector pairs overlaid onto the HAADF image in **d**.

## 4. Ti *a-b* Plane Displacements

All the methods discussed so far provide evidence for Ti *a-b* plane displacements, but the nature of its order is unclear. To check the local ordering, if any, of these displacements, we carried out electron diffraction and imaging studies. Figure 2c shows a diffraction pattern viewed along the [1$\bar{1}$0]-zone axis obtained using a transmission electron microscope (TEM). Between the Bragg spots, we observe diffuse scattering rods (highlighted by blue arrows) that are specifically oriented along the 110-direction in reciprocal space (more details in Supporting Information Section IV). These streaks suggest the presence of strain or disorder normal to the (110)-planes. We do not observe any streaks between the Bragg reflections along 001. Diffuse rods were also observed along 100 in Figure S5d but are weaker in intensity. These diffraction patterns suggest that the observed Ti *a-b* plane displacements could possess short-range ordering.

To directly visualize the local ordering of these *a-b* plane displacements, we performed atomic scale imaging using an aberration-corrected scanning transmission electron microscope (STEM). Figure 2d shows a high-angle annular dark field (HAADF) STEM image of BaTiS$_3$ viewed along the [001]-zone axis. In this imaging mode, the intensity is proportional to the square of the average atomic number ($Z^2$) of the columns[31]. Thus, the heavier Ba atomic columns appear brighter than the lighter Ti columns, while the lightest S columns are almost invisible due to the dynamic-range constraints of the detector. We extracted the position of the atomic columns by fitting 2D Gaussians as shown in the small field-of-view HAADF images of three representative regions in Figure 2e. We observe that the Ti atomic columns are displaced away from the centroid position defined by the six adjacent Ba columns. The Ti off-center displacements along the *a-b* plane have been extracted for the entire region shown in Figure 2d and are overlaid as displacement vectors on a vector-color map in Figure 2f. While most of the Ti atomic columns are displaced away from the centroid, some Ti atomic columns have nearly zero displacements and are largely present at the boundaries of the domains having columns with displacements along different orientations. Figure 2g shows a histogram of the magnitude of Ti off-center displacements obtained from Figure 2f. By fitting a Gaussian curve to the histogram, the average displacement of Ti atomic columns along the *a-b* plane is derived to be 0.11 Å. The average Ti displacement measured from STEM is in good agreement with SC-XRD refinement, where *a-b* plane displacements of $TiS_6$-chain A and chain B are 0.104 Å and 0.142 Å [Table S6], with an average displacement of 0.129 Å.

It is apparent from the displacement vector color map shown in Figure 2f that the orientation of Ti displacements in neighboring unit cells are somewhat aligned giving rise to a domain-like structure, while displacements farther away from any unit cell are randomly oriented. To quantify the correlation between the orientation of the Ti displacements, we calculated the Pearson correlation coefficient (Pearson's *r*)[32], which evaluates the linear correlation between any pair of vectors with a value varying from -1 to 1, where 0 indicates no correlation (disorder), 1 shows a perfect correlation, i.e., the vectors are parallel, and −1 signifies a negative correlation, i.e., the vectors are antiparallel. (See details in the Supporting Information Section V). The distribution of the short-range correlation between every Ti atomic column with its six nearest neighboring columns is shown as a color map in Figure 2h, with red, blue, and white representing positive, negative, and zero correlation, respectively, to the neighboring cells. A histogram of the linear short-range correlations for the entire region in Figure 2d shows that most of Ti *a-b* plane displacements have a strong positive correlation with their nearest neighbors with a mean value of ~0.6 [Figure S7c]. In contrast, the long-range correlation obtained by averaging the correlation between any Ti column displacement with all the other displacements in the entire region shown in Figure 2d, shows no meaningful correlation [Figures S7d and S7e], in support of the *P*6$_3$*cm* space group observed from the macroscale XRD measurements.

## 5. Giant Refractive Index Modulation

To gain insights into the effect of the Ti *a-b* plane displacements on the optical properties, we performed first-principles DFT calculations[33]. As mentioned before, the *P*6$_3$*cm* structure does not have any *a-b* plane displacement of Ti atoms. A group symmetry analysis shows that freezing the $\Gamma_5$ distortion mode in the *P*6$_3$*cm* structure leads to the off-centering of Ti atoms along the *a-b* plane. These displacements are ordered in a pattern shown in Figure S10, with neighboring Ti atoms aligned antiparallel. Freezing the $\Gamma_5$ distortion mode lowers the energy of the system, as shown in Figure 3a, which explains the presence of *a-b* plane displacements in the experiments. We then calculated the complex dielectric function, $(n + i\kappa)$, of BaTiS$_3$ with different amplitudes of the $\Gamma_5$ distortion mode frozen to the *P*6$_3$*cm* structure [Figure S11 and S12]. The dielectric function was calculated along the *c*-axis, which we refer to as the extraordinary axis, and perpendicular to it, which are the ordinary axes. The real parts of the ordinary ($n_o$) and the extraordinary ($n_e$) refractive indices calculated in the transparent range of BaTiS$_3$ for photon energies smaller than 0.3 eV (light wavelength > 4 μm); and their difference, which is the birefringence ($\Delta n = n_e - n_o$), as a function of the average Ti displacement in the *a-b* plane is presented in the middle and bottom panels of Figure 3a, respectively. We find that $\Delta n$ increases with increasing Ti *a-b* plane displacements until it plateaus for displacements > 0.3 Å, as shown in the bottom panel of Figure 3a. We also find that $n_e$ remains almost unchanged with increasing Ti *a-b* plane displacements. So, the dominant contribution to the increase in $\Delta n$ comes from a decrease in the magnitude of $n_o$ with increasing Ti *a-b* plane displacements. These trends are also observed for the wavelength-dependent $\Delta n$ [Figure S11b] and dichroism ($\Delta \kappa = \kappa_e - \kappa_o$), [Figure S11a] with varying displacements.

To identify the electronic origin of the enhancement in $\Delta n$ with Ti *a-b* plane displacements, we investigated the evolution of the occupied electronic states. $\Delta n$ in a crystal arises from the distribution and orientation of valence electrons near the Fermi energy.[21,34] Thus, electron redistribution near the Fermi energy introduced by the *a-b* plane Ti displacements can be expected to change the anisotropic optical response of BaTiS$_3$. The valence band is primarily constituted by S-3*p* states, as confirmed by the atom- and orbital-projected density of states (PDOS) shown in Figure S15a. We find that within 0.5 eV of the Fermi energy, the S-3*p* states have a non-bonding character, as shown in Figure S15b, and are most likely to be polarized.[34]

Ti displacements in the *a-b* plane result in a decrease in the density of electrons that are oriented along the *a-b* plane and located within 0.5 eV below the Fermi energy, as shown in the integrated electron density plots in Figure 3b. These electrons primarily belong to S-$3p_x$ and $3p_y$ orbitals. The decreasing occupation of S-$3p_x$ and $3p_y$ states with increasing Ti *a-b* plane displacements is further confirmed from the PDOS integrated within -0.5 eV to the Fermi energy (set to 0 eV), as shown in Figure S17b. The decreasing electron distribution within the *a-b* plane has a direct correlation with the decreasing $n_o$ with *a-b* plane Ti displacements, as shown in Figure S17b, d.

We also performed spectral weight sum rule analysis for BaTiS$_3$ with different Ti *a-b* plane displacements (See details in Figures S18 and S19 in Supporting Information Section VII). As the Ti *a-b* plane displacements increase, the electron density contributing to the interaction with the electric field polarized along the *a-b* plane shows a progressive decrease as a function of energy (frequency) as shown in Figure S18a. On the contrary, we found a slight increase of the electron density contributing to the dielectric response along the *c*-axis in Figure S18b. These opposite trends align well with the spatial distribution of electron density, as illustrated in Figure 3b and Figure S17, wherein the density of electrons — that are oriented along the *a-b* plane (S-$3p_x$ and $3p_y$ states) and located within 0.5 eV below the Fermi energy — decreases with increasing Ti *a-b* plane displacements.

Finally, we have investigated the effect of the experimentally observed disordered Ti *a-b* plane displacements on $\Delta n$, as opposed to the ordered Ti displacements simulated by freezing the $\varGamma_5$ distortion mode. We performed *ab initio* molecular dynamics (AIMD) simulations starting with BaTiS$_3$ with ordered $\varGamma_5$ distortions and equilibrated the structure at 800 K for 5 ps to randomize the Ti displacements [Figure S14]. We then used 8 randomly selected snapshots and calculated their dielectric function, $\Delta n$ and $\Delta \kappa$. The average values and standard deviations (Std.) of the calculated $\Delta n$ and $\Delta \kappa$ as a function of wavelength are shown in Figure 3c. These results are in excellent agreement with the experimental results, which show that the hybridization of Ti and S states is, by large, determined by the local octahedral distortions, and is less sensitive to the Ti and S atoms in adjacent TiS$_6$-chains. Furthermore, we observe that the magnitude of Ti *a-b* plane displacements correlates extremely well with charge redistribution and the optical anisotropy in BaTiS$_3$.

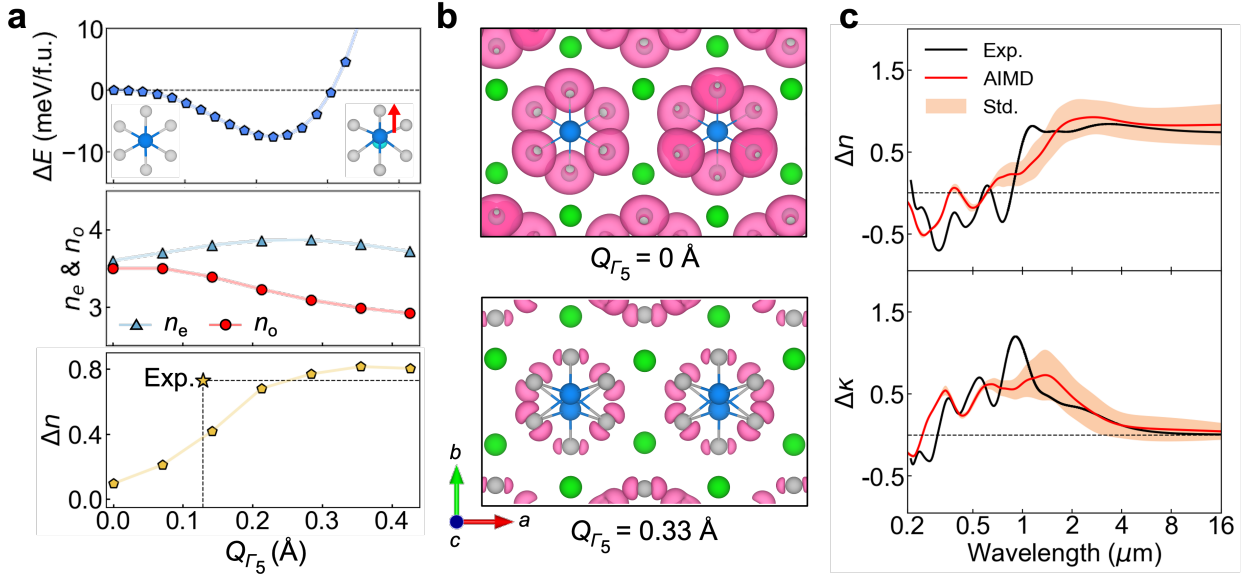

**Figure 3**. **Optical anisotropy enhanced by Ti *a-b* plane displacements**. **a**, Top panel: Energy of BaTiS$_3$ as a function of the $\Gamma_5$ distortion mode associated with ordered *a-b* plane displacements of Ti, as shown in atomic models in the inset. Middle and bottom panels: Real part of refractive index along the ordinary ($n_o$) and extraordinary ($n_e$) axes, and birefringence ($\Delta n = n_e - n_o$) of BaTiS$_3$ (averaged between 4-16 μm wavelengths) as a function of average Ti *a-b* plane displacements. **b,** Spatial distribution of electrons within 0.5 eV below the Fermi energy projected onto the *a-b* plane (top panel) without and (bottom panel) with Ti *a-b* plane displacements. Ti *a-b* plane displacements significantly decrease the charge distribution within the *a-b* plane, which is the reason for the decrease in $n_o$ shown in the middle panel of **a**. Isosurface in **b** set to an electron density of 0.0025 e/Å$^3$. **c,** Birefringence ($\Delta n$) and dichroism spectra ($\Delta \kappa$) of the structure obtained by averaging 8 snapshots from an *ab*-initio molecular dynamics run of BaTiS$_3$ equilibrated at 800 K. The averaged structure has randomized Ti *a-b* plane displacements. The standard deviation (Std.) of the 8 structures is shaded in orange and centered around the average shown by the red solid line. The black solid line shows the experimentally measured anisotropy.

## 5. Conclusions

BaTiS$_3$, at room temperature, shows correlated disorder. Off-center Ti displacements within the *a-b* plane have long-range disorder but are ordered along the *c*-axis. The ordinary ($n_o$) refractive index in the partial (1.5 – 5 μm) and fully transparent region (> 5 μm) decreases while the extraordinary refractive index ($n_e$) remains almost unchanged with increasing Ti *a-b* plane displacements [Figure S12b, d]. The overall effect is a large increase of ~0.8 in the birefringence, $\Delta n$. Our results suggest that the refractive index tensor of BaTiS$_3$ is highly sensitive to the magnitude of the Ti displacements, regardless of whether the displacements are ordered or disordered. The exact effect of the correlated disorder of the Ti-sublattice, as observed experimentally, on the optical properties remains an open question. The magnitude of Ti displacements can potentially be tuned *via* external stimuli, such as temperature, strain, or electric field, in a manner of "ferroic" switching between different correlated Ti displacement modes. Therefore, the potential for reversible symmetry breaking by *a-b* plane Ti displacements in the $TiS_6$ octahedra in $BaTiS_3$ and related perovskite chalcogenides makes them a good platform to achieve tunable anisotropy, large non-linearity, and coupled phenomena such as opto-elastic and electro-optic effects. The adjustable refractive indices in $BaTiS_3$ can facilitate miniaturized controlled retarders and variable waveplates for polarization manipulation, and electro-optic modulators towards a broad range of mid-infrared photonic applications.

## 6. Methods

*$BaTiS_3$ Crystal Growth*: Experimental Details. $BaTiS_3$ crystallizes on top of polycrystalline $BaTiS_3$ by chemical vapor transport in a sealed quartz ampule with iodine as a transporting agent[12,19]. Starting materials, barium sulfide powder (Sigma-Aldrich, 99.9%), titanium powder (Alfa Aesar, 99.9%), sulfur pieces (Alfa Aesar, 99.999%), and iodine pieces (Alfa Aesar 99.99%) were stored and handled in a nitrogen-filled glove box. Stoichiometric quantities of precursor powders with a total weight of 1.0 g were mixed and loaded into a quartz tube of 15 mm inner diameter along with around 0.75 mg/cm$^3$ iodine inside the glove box. The tube was capped with ultra-torr fittings and a bonnet needle valve to be evacuated and sealed using a blowtorch to avoid exposure to air. The sealed tube was about 12 cm in length and was loaded and heated to 1050 °C at 100 °C/h and held for 100 h before a slow cooling down to 950 °C at 10 °C/h in a Lindberg/Blue M Mini-Mite Tube Furnace. The furnace was then shut off letting the ampule cool down within the furnace.

*Synchrotron Single Crystal Diffraction*: Single crystal diffraction at room temperature was carried out on beamline 12.2.1 at the Advanced Light Source, Lawrence Berkeley National Laboratory. Crystals were mounted on MiTeGen Kapton loops and placed in a nitrogen cold stream on the goniometer head of a Bruker D8 diffractometer, which is equipped with a PHOTONII CPAD detector operating in shutter-less mode. Diffraction data were collected using synchrotron radiation monochromate with a wavelength of 0.72880 Å with silicon (111). A combination of *f* and *ω* scans with scan speeds of 0.25 s per degree for the *f* scans, and 10 s per degree for the ω scans at 2θ = 0 and -20°, respectively, to extract data for strong reflections.

On top of that, weak reflections are collected again by repeating the ω scans under stronger incident synchrotron radiation. Unit cell determination, integration, and scaling are then carried out in Bruker APEX 3, where the precession maps of 1.5 Å resolution were generated under the refined unit cell. Crystal structure refinement and Fourier difference map analysis were done in ShelXle[35]. Detailed data collection approach, crystallography analysis, and refinement results at different temperatures are listed in Supporting Information Section VIII.

*Scanning Transmission Electron Microscopy*: The electron-transparent lamellae were prepared using Ar-ion milling (Fischione Model 1010) with a liquid-nitrogen cooling stage for both [001] and [100] orientated $BaTiS_3$ samples. Atomic-resolution STEM imaging was performed using the aberration-corrected Nion UltraSTEM™ 100 at CNMS, ORNL, operated at 100 kV with a convergence angle of 30 mrad. *Z*-contrast HAADF-STEM images were acquired using an annular dark field detector with a collection angle of 80–200 mrad. The selected area electron diffraction patterns were acquired using a JEOL JEM-2100F Field Emission (S)TEM at IMSE, WUSTL, operated at 200 kV.

Atomic column positions in the HAADF images were determined using a 2D Gaussian peak-fitting algorithm. The Ba and Ti atomic columns were indexed according to peak intensity. The off-centering displacements of Ti atoms were then measured as the offset of Ti relative to the centroid of the hexagon formed by its nearest neighboring Ba atoms.

*First-Principles Calculations*: DFT calculations were performed using the Vienna Ab initio Simulation Package (VASP)[29,30]. We used projector augmented-wave (PAW) potentials and the generalized gradient approximation within the Perdew-Burke-Ernzerhof (GGA-PBE) parameterization to describe the electron-ion and the electronic exchange-correlation interactions[36–38], respectively. The PAW potentials explicitly included $3s^23p^4$ for S, $3p^63d^24s^2$ for Ti, and $5s^25p^66s^2$ for Ba as valence electrons. An energy cutoff of 650 eV has been used to truncate the plane-wave basis set according to convergence tests. The Brillouin zone was sampled using a $\Gamma$-centered *k*-points mesh with a spacing of 0.025 Å$^{-1}$. Convergence was reached with energy change under $10^{-7}$ eV; ionic positions and cell constants were fully relaxed until the forces were smaller than $10^{-3}$ eV/Å. Spin-polarization was included in all our calculations.

The distortion modes of Ti displacement from a higher symmetry $BaTiS_3$ structure were investigated using the ISODISTORT software suite[39]. To reduce the computational cost, *Ab initio* molecular dynamics (AIMD) was used to thermalize a 2 × 2 × 1 *P6₃cm*-$BaTiS_3$ supercell to 800 K. A Γ-point sampling of *k*-points was used with an MD time step of 0.5 fs. Langevin thermostat was used to control temperature[40]. Microcanonical *NPT* dynamics were run for 5 ps to equilibrate the $BaTiS_3$ structure at 800 K. 8 snapshots of the equilibrated $BaTiS_3$ were used for optical property calculations. The frequency-dependent dielectric function along the different crystal axes was calculated by DFT methods with the LOPTICS tag in the independent particle approximation (IPA)[33] as implemented within VASP. The chemical bonding analysis was performed using crystal orbital Hamilton population (COHP) as implemented in the LOBSTER package[41].

*Solid-state $^{47,49}$Ti NMR*: $^{47,49}$Ti ssNMR spectra were recorded at $B_0$ = 18.8 T (800 MHz for $^1$H) using a standard-bore Bruker BioSpin spectrometer equipped with an AVANCE-III console and a 3.2 mm HX MAS probe tuned to $^{49}$Ti at 47.120 MHz. Samples were packed in a $ZrO_2$ rotor and closed with Vespel® caps. NMR spectra were acquired under static conditions and $^{49}$Ti chemical shifts were externally referenced against pure Ti-isopropoxide at −850 ppm (corresponding to $TiCl_4$ at 0 ppm).

$^{47,49}$Ti NMR spectra were acquired using the WURST-QCPMG experiment described by O'Dell and Schurko[42]. In this experiment, Wideband Uniform-Rate Smooth Truncation (WURST) pulses were used to circumvent the limited excitation bandwidths provided by ordinary rectangular pulses, while the Quadrupolar Carr–Purcell–Meiboom–Gill (QCPMG) sequence was used as a signal enhancement procedure, whereby the signal intensity is concentrated in evenly spaced spikelets. An 8-step phase cycle was used to select both $p$ = +1 and −1 coherence pathways, and the delay times were adjusted so that the echoes from each pathway were coincident.

The first excitation pulse (WURST-A) consisted of a 50 μs symmetric WURST 80 pulse whose frequency was swept adiabatically from positive to negative offset frequencies at a constant rate. This excitation WURST-80 pulse resulted in a 750 kHz excitation bandwidth and its power level was optimized experimentally on $BaTiO_3$[43]. The second refocusing pulse (WURST-B) used in the QCPMG procedure was identical to WURST-A, yielding frequency-dispersed echoes. No compensation of the frequency-dispersed echoes was carried out, leading to a slight distortion of the baseline.

For $BaTiS_3$, the sample temperature was regulated at 270 K to improve the signal-to-noise ratio. 50 full echoes were acquired and averaged over 655,360 transients using a recycle delay of 0.1 s, leading to an overall acquisition time of 23 hours. The WURST–QCPMG spectrum was processed with 100 Hz of exponential apodization, followed by magnitude calculation after Fourier transformation to obtain absorptive spikelets using the Topspin 3.6 software package. The resulting $^{47,49}Ti$ NMR spectrum was fitted using a special build of DMfit software provided by Dr. D. Massiot[44].

*First principles calculations of $^{47}Ti$ NMR parameters*: $^{47}Ti$ NMR parameters were calculated using the CASTEP[45] software package. The generalized gradient approximation by Perdew, Burke, and Ernzerhof[36] was employed to approximate the exchange-correlation term. All calculations used "on-the-fly" ultrasoft pseudopotentials as supplied by CASTEP. NMR calculations were performed using the projector augmented-wave method (GIPAW)[46,47]. The scalar-relativistic zeroth-order regular approximation (ZORA[48]) was employed to account for relativistic effects. As input structure, we either used the $BaTiS_3$ experimental structure obtained from single crystal X-ray diffraction, or a $BaTiS_3$ structure optimized from first principles using the VASP software package. The procedure for each calculation involved the convergence of chemical shielding constants with respect to plane-wave cutoff energy and *k*-point grid density. Converged cutoff energies and *k* meshes are listed in Table S1.

**Supporting Information**

Supporting Information is available from the Wiley Online Library or from the author.

**Acknowledgments**

This work was supported by the Army Research Office (ARO) under award number W911NF-19-1-0137 and via an ARO MURI program with award number W911NF-21-1-0327, the National Science Foundation under grant numbers DMR-2122070, 2122071, and 2145797, an Air Force Office of Scientific Research grant no. FA9550-22-1-0117, and the USC Provost New Strategic Directions for Research Award. The ssNMR research is partially supported by the UCSB NSF Quantum Foundry through Q-AMASE-i program award number DMR-1906325. The synchrotron research used resources of the Advanced Light Source, which is a DOE Office of Science User Facility under contract no. DE-AC02-05CH11231. Electron microscopy was supported by the U.S. Department of Energy (DOE), Office of Science, Basic Energy Sciences, Materials Sciences and Engineering Division and conducted at the Center for Nanophase Materials Sciences (CNMS), which is a US Department of Energy, Office of Science User Facility at Oak Ridge National Laboratory. Neutron scattering work by M. E. M. and R. P. H. supported by the US Department of Energy (DOE), Office of Science, Office of Basic Energy Sciences (BES), Materials Sciences and Engineering Division. A portion of this research used resources at the Spallation Neutron Source, supported by DOE, BES, Scientific User Facilities Division. This work used computational resources through allocation DMR160007 from the Advanced Cyberinfrastructure Coordination Ecosystem: Services & Support (ACCESS) program, which is supported by NSF grants # 2138259, #2138286, #2138307, #2137603, and #2138296. H.M. and M.K. acknowledge the support from the Office of Naval Research (N00014-20-1-2297). The authors gratefully acknowledge the use of facilities at the Core Center for Excellence in Nano Imaging at the University of Southern California. Dr. Dominique Massiot is gratefully acknowledged for providing a special build version of the DMfit software and useful discussions on $^{47,49}$Ti ssNMR. B.Z acknowledges technical assistance from Mythili Surendran, Thomas Orvis, and Harish Kumarasubramanian in collaboration with the related projects.

**Author Contributions**

B.Z. and G.R. contributed equally to this work. B.Z., S.S., H.C., S.N., and J.R. developed, optimized, and synthesized the single crystals of $BaTiS_3$. G.R., A.S.T., A.R.L., M.C., J.A.H., and R.M. carried out the TEM and STEM measurements and their analyses. H.M., J.S., and M.A.K. verified the optical properties of $BaTiS_3$ and discovered the inconsistency between experiment and theory. V.C.W., R.G., and R.J.C. worked out the ssNMR measurement and simulations of $BaTiS_3$. G.R., G.-Y.J., and R.M. performed the first-principles DFT calculations of $BaTiS_3$. B.Z., N.S., B.C.C., S.J.T., and J.R. carried out SC-XRD and refined the $BaTiS_3$ structure in detail. B.Z., R.P.H., M.E.M., and J.R. reanalyzed the neutron scattering PDF. B.Z., G.R., J.R., and R.M. prepared the manuscript with edits from all the authors.

**Competing Interests**

The authors declare no conflict of interest.

**Data Availability**

Data presented in the main text and supporting information are open access and can be found on a Zenodo repository (https://zenodo.org/doi/10.5281/zenodo.8415411).

*Notice:* This manuscript has been authored by UT-Battelle, LLC, under Contract No. DE-AC0500OR22725 with the U.S. Department of Energy. The United States Government retains and the publisher, by accepting the article for publication, acknowledges that the United States Government retains a non-exclusive, paid-up, irrevocable, worldwide license to publish or reproduce the published form of this manuscript or allow others to do so, for the United States Government purposes. The Department of Energy will provide public access to these results of federally sponsored research in accordance with the DOE Public Access Plan (http://energy.gov/downloads/doe-public-access-plan).

At room temperature, quasi-one-dimensional chalcogenide crystal $BaTiS_3$ shows structural disorder due to off-centering Ti displacements. These disordered Ti displacements can selectively reduce the refractive index along the *ab*-plane while having minimal effect on the refractive index along the chain direction, which subsequently results in a giant enhancement of optical anisotropy. The disordered Ti displacements can offer another degree of freedom for light manipulation in photonic applications.

*Boyang Zhao, Guodong Ren, Hongyan Mei, Vincent C. Wu, Shantanu Singh,*
*Gwan-Yeong Jung, Huandong Chen, Raynald Giovine, Shanyuan Niu, Arashdeep S. Thind,*
*Jad Salman, Nick S. Settineri, Bryan C. Chakoumakos, Michael E. Manley,*
*Raphael P. Hermann, Andrew R. Lupini, Miaofang Chi, Jordan A. Hachtel, Arkadiy Simonov,*
*Simon J. Teat, Raphaële J. Clément, Mikhail A. Kats, Jayakanth Ravichandran, and*
*Rohan Mishra*

**Giant Modulation of Refractive Index from Picoscale Atomic Displacements**

ToC figure ((Please choose one size: 55 mm broad × 50 mm high **or** 110 mm broad × 20 mm high.  Please do not use any other dimensions))

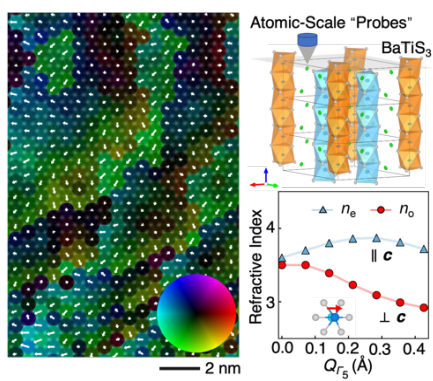